\documentclass[preprint]{revtex4-1}
\usepackage{graphicx}
\usepackage{bm}
\usepackage{wasysym}
\usepackage{setspace}
\usepackage{mathrsfs}

\begin{document}

%\preprint{\sf Version 1 (\today)}
\title{Lattice Theoretical Approach in Strongly Correlated Electron Systems}
\author{Mu-Kun Lee}
\author{Chyh-Hong Chern}
\email{chchern@ntu.edu.tw} 
\email{chern@alumni.stanford.edu}
\affiliation{Department of Physics, National Taiwan University, Taipei 10617, Taiwan}
\date{\today}
\begin{abstract}
The effective lattice models in strongly correlated electron systems are \emph{derived} in particular for the cuprate superconductors, that incorporate the quantum fluctuations of the spin Berry's phase and the antiferromagnetic fluctuation.  Consistent with the field-theoretical approach, the density modulation, the weak ferromagnetism, and the superconductivity are reproduced.  We discussed the pros and cons of the effective-model approach and demonstrate that both positive and negative Hubbard model are the effective models subject to the occurrence of the quantum fluctuations of the correlation degrees of freedom.
\end{abstract}

%\pacs{Valid PACS appear here}
\maketitle
\noindent
\emph{Introduction}

Effective lattice modeling is necessary in condensed matter physics, since the Grand Hamiltonian~\cite{laughlin2000}, namely the Coulomb interaction between electrons, is not exactly solvable.  In this approach, the only guiding principle of the model building is the experimental facts.  In the strongly correlated electron systems, we have encountered tremendous failures in this approach, especially in cuprate superconductors, since the exotic experimental facts, that go beyond our knowledges, keep appearing.  Among all of them, the most exotic ones are arguably the pseudogap formation~\cite{shen2014}, the Fermi-arc formation, weak ferromagnetic time-reversal-symmetry breaking~\cite{kapitulnik2008}, charge density modulations~\cite{julien2011, keimer2012}, and the superconductivity~\cite{bednorz1986}.  The physics of cuprate superconductors are thought to be difficult, as well as other correlated electron systems, because there is no single model that explains all experiments.  In this paper, we shall show that the above statement is true.  While the positive Hubbard model describes the uniform state, the Hubbard term has to change sign to describe the inhomogeneous state in the derivation of the effective models.  Furthermore, the systematic method to study correlated electrons will be outlined.

The correlation is a pure quantum-mechanical concept that is the consequence of the wavefunction overlaps.  It is a non-perturbative effect and generates dynamics additionally.  For example, the antiferromagnetic correlation is the consequence of the spatial wavefunction overlaps.  The overlap of the spin wavefunction gives rise to the spin-Berry's phase,
\begin{eqnarray}
\vec{a}=-iz^\dagger_\alpha\vec{\nabla}z_\alpha,
\end{eqnarray}
where $z_\alpha$ are the spin wavefunctions, and $z^\dagger_\alpha z_\alpha=1$.  The spin-Berry's phase is a gauge field in nature~\cite{wen1989}.  Namely, taking $z_\alpha \rightarrow e^{i\lambda(x)} z_\alpha$, $\vec{a} \rightarrow \vec{a}+\vec{\nabla}\lambda$ defines the gauge transformation.  As an alternative approach, the correlations are considered as the explicit degrees of freedom that give rise to the dynamics to the electrons, and the effective models can be derived.  Assuming that the anti-ferromagnetic correlation and the spin-Berry's phase are the significant correlations in the cuprate superconductors, a possible dynamics to the electrons has been proposed by one of us (CHC)~\cite{chern2014},
\begin{eqnarray}
\mathscr{L} = \sum_{\sigma}\Big\{&\psi^\dag_{\sigma}&(i\partial_0)\psi_\sigma-\frac{1}{2m}\Big[(-\frac{\vec{\nabla}}{i}-g\vec{a})\psi^\dag_\sigma\Big]\Big[(\frac{\vec{\nabla}}{i}-g\vec{a})\psi_\sigma\Big] -ga_0\psi^\dag_\sigma\psi_\sigma\Big\}-\frac{1}{4}f_{\mu\nu}f_{\mu\nu}\nonumber \\ &+&\frac{1}{2}M_0^2(D_0 \phi)^\dag(D_0 \phi)-\frac{1}{2}M_1^2(D_i \phi)^\dag(D_i \phi),\label{fulllagrangian}\end{eqnarray}
where the antiferromagnetic correlation is parametrized by $\phi (t, \vec{x})= \frac{1}{q}e^{i\sigma(t, \vec{x})}$, $q$ is its coupling to the spin-Berry's phase, and $D_0 = i\partial_0 - qa_0$ and $D_i = -i\partial_i - qa_i$, and $\psi_\sigma$ are the electron coordinates.  Integrating over the degrees of freedom of the correlations, the effective electronic Lagrangian in the pseudogap phase was computed~\cite{chern2014}
\begin{eqnarray}
\mathscr{L}_{\text{eff}}= \sum_{\sigma}&\psi^\dag_{\sigma}&(x) (i\partial_0 +\frac{\nabla^2}{2m})\psi_{\sigma}(x) +\frac{-ig^2}{2}\sum_{\sigma,\sigma'}\rho_{\sigma}(x)\frac{i}{k^2-M_0^2+i\eta}\rho_{\sigma'}(x)\nonumber\\&-&\frac{-ig^2}{2}\sum_{\sigma,\sigma'}\vec{J}_{\sigma}(x)\frac{i}{k^2-M_1^2+i\eta}\cdot \vec{J}_{\sigma'}(x), \label{lagrangian2}
\end{eqnarray}
where $k^2=\omega^2-|\vec{k}|^2$, and $\rho_{\sigma}$ and $\vec{J}_{\sigma}$ are the density and the current operators respectively.  In the $|\vec{k}|\rightarrow 0$ limit, Eq.~(\ref{lagrangian2}) reduces to the positive Hubbard model.  The contact repulsive interaction is the cause of the pseudogap formation~\cite{chern2014}.

The electronic structure in cuprate superconductors is not homogeneous.  The Fermi-arc formation~\cite{shen2014}, the charge density modulation~\cite{julien2011, keimer2012}, and the  ferromagnetic time-reversal-symmetry breaking~\cite{kapitulnik2008} robustly exist.   The inhomogeneity suggests that we should look into the $\frac{k^2}{M^2_{0,1}}$ expansions in the Eq.~(\ref{lagrangian2}) for the effective Hamiltonian.  Actually, those phenomena have been shown to be the consequences of the quantum fluctuations of the antiferromagnetic correlation and the spin-Berry's phase~\cite{Lee2018}.  In this paper, we will show that those results can be reproduced in the $\frac{k^2}{M^2_{0,1}}$ expansions.  In the previous approach, the dynamics of electrons due to the quantum fluctuations was discussed only at the classical level~\cite{Lee2018}.  Namely, the quantum fluctuation is taken as the external force applying to the electrons by the Newton's laws.   The current effective-model approach advances to the quantum dynamics, that may as well facilitates the numerical computation.

Before deriving the effective model, let us comment about the quantum fluctuations.  The concept of the quantum fluctuations has been applied to all fields in physics.  In the correlated electron systems, quantum fluctuations are often discussed in the context of quantum criticality.  However, we should emphasize that the quantum fluctuations in cuprate superconductors have nothing to do with the quantum criticality~\cite{Lee2018}.  They are the quantum fluctuations of the gauge degrees of freedom in the paramagnetic correlated electronic phase.

\vspace{0.5cm}
\noindent
{\it Derivation}

\begin{figure}[htb]
\includegraphics[width=0.45\textwidth]{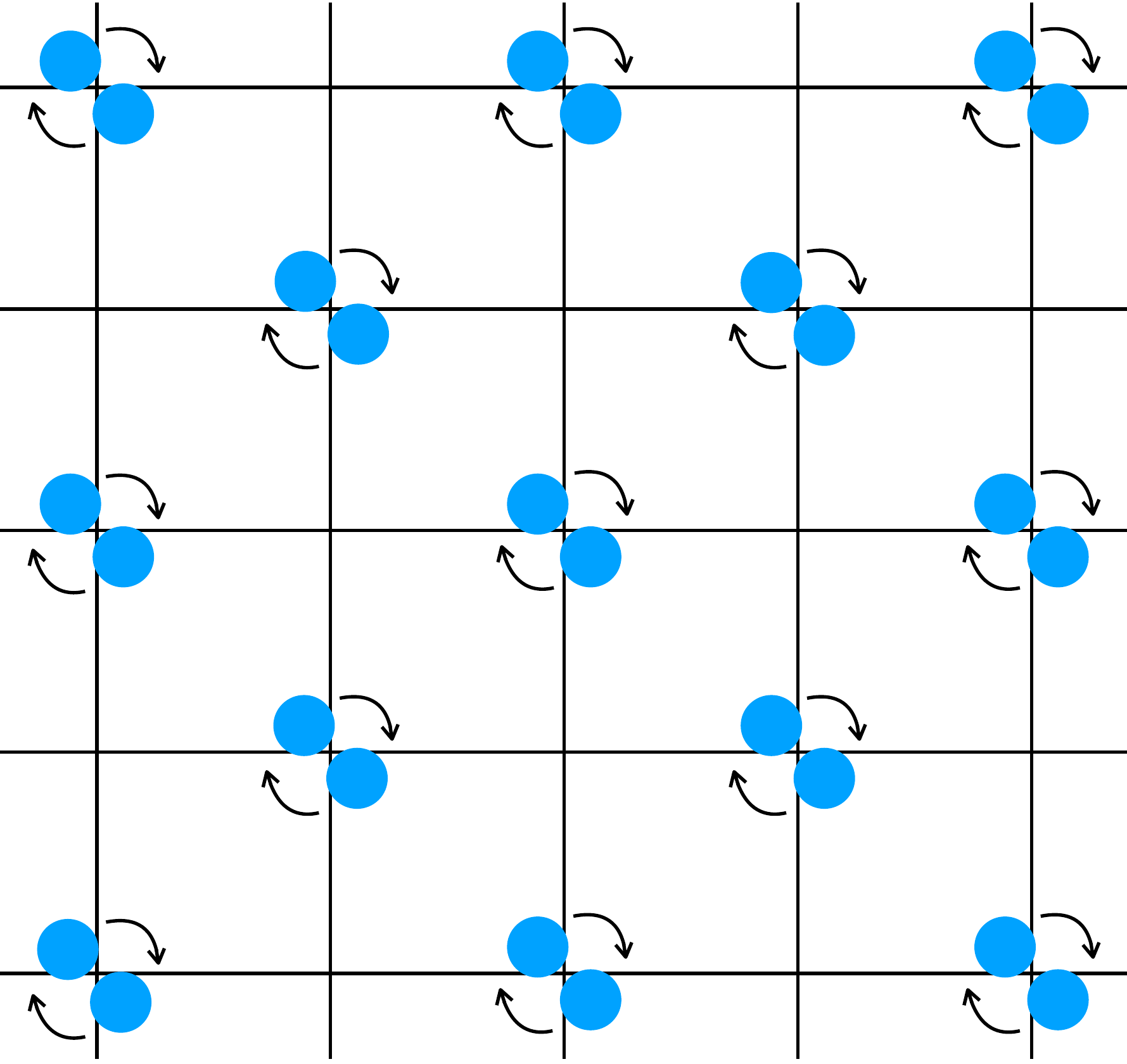}
\caption{A ground state configuration of Eq.~(\ref{LCmodel}).  For the cuprates, the vertices of the lattice denote the copper sites.  The blue balls denote the electrons.  The arrows denote the possible rotational motion responsible for the polar Kerr rotation signals.}\label{cdw}
\end{figure}

Let us expand the interaction terms in Eq.~(\ref{lagrangian2}).  The density-density interaction is given by
\begin{eqnarray}
\frac{ig^2}{2}\sum_{\sigma,\sigma'}\rho_\sigma(x)\frac{i}{k^2-M^2_0+i\eta}\rho_{\sigma'}(x)\approx\frac{g^2}{2M^2_0}\sum_{\sigma,\sigma'}\rho_\sigma(x)\rho_{\sigma'}(x)+\frac{g^2}{2M^2_0}\sum_{\sigma,\sigma'}\rho_\sigma(x)\frac{k^2}{M^2_0}\rho_{\sigma'}(x). \label{e1}
\end{eqnarray} 
The first term in Eq.~(\ref{e1}) is positive indicating the repulsive interaction and corresponds to the Hubbard on-site interaction, which is the origin of the pseudogap formation in the homogeneous state~\cite{chern2014}.  The second term in Eq.~(\ref{e1}) is the interaction energy for the inhomogeneous states, that is given by
\begin{eqnarray}
\frac{g^2}{2M^2_0}\sum_{\sigma,\sigma'}\rho_\sigma(x)\frac{(-\partial^2_t+\nabla^2)}{M^2_0}\rho_{\sigma'}(x). \label{e2}
\end{eqnarray}
The first term in Eq.~(\ref{e2}) is related to the energy of the inhomogeneous state, that is determined by the onset temperature of the charge density modulation $\sim (\frac{k_BT_{\text{CDW}}}{M_0})^2\sim 10^{-8}$ and can be neglected~\cite{Lee2018}.  Discretizing the second term, we obtain
\begin{eqnarray}
\sum_{\sigma\sigma'}\rho_{\sigma}(x_i)\frac{\nabla^2}{M_0^2}\rho_{\sigma'}(x_i)=-\frac{4}{M_0^2c^2_0}\sum_{\sigma\sigma'}\rho_{\sigma}(x_i)\rho_{\sigma'}(x_i)+\frac{1}{M_0^2c^2_0}\sum_{j=i\pm\hat{e}_x, i\pm\hat{e}_y \atop \sigma\sigma'}\rho_{\sigma}(x_i)\rho_{\sigma'}(x_j), \label{e3}
\end{eqnarray} 
containing the contact interaction and the interaction between the nearest neighbor sites.  The first term in Eq.~(\ref{e3}) is negative, and the second term is positive.  The on-site attractive interaction pulls the electrons together to stay on the same site.  Together with the repulsive nearest neighbor interaction, the inhomogeneous state favors a checkerboard pattern of the density modulation as shown in Fig.~(\ref{cdw}), that is consistent with our previous work and the experiments~\cite{julien2011, chern2018}.  Due to the Pauli exclusion principle, two electrons on the same site must have opposite spin orientations.  The mobility of the electron pairs enhances as the system is doped with the holes.  As the temperature is low, the electrons on the same site form the superconducting pair, and the superconductivity arises.

It is observed that the length scale of the charge density modulation is the lattice scale, where the wavelength is roughly $4c_0$ in cuprates~\cite{julien2011, keimer2012, davis2014, davis2016a, davis2016b}.  $M_0$, defining the length scale of the interaction, is in the same scale of the density modulation.  Namely, $M_0c_0$ is of order of unity, which indicates that the contact interaction in Eq.~(\ref{e3}) modifies the repulsive Hubbard term to be attractive.  This result echoes our mechanism of the superconductivity in cuprates~\cite{chern2018}, where the presence of the quantum fluctuations of the antiferromagnetic correlation effectively generates the attractive force to cancel the repulsive force that leads to pseudogap formation.  The attractive force pairs the electrons up in the real space without a Fermi surface in distinction to the Cooper pairs.  However, the effective model approach does not contain full informations of the quantum fluctuations.  Firstly, it does not pick up the phase information of the quantum fluctuations, where the effective potentials can be of $d$-type or $s$-type.   Secondly, the effect of the quantum fluctuations is an external source similar to the wind.  The density modulation is the electronic ripples blown by the quantum fluctuations.  As the wind stops, ripples disappear, that does not involve any phase transition.  

Similar to all situations requiring the analysis of Taylor expansion, the problem of the convergence needs to be carefully treated.  In our case, $k\sim \frac{1}{c_0}$ and $\frac{k}{M_0}\sim O(1)$.  The infinite expansion is certainly divergent.  However, the essence of the effective model approach is to extract physics in the \emph{correct} scale.  Roughly speaking, the $(\frac{k^{2}}{M_0^{2}})^n$ term in the expansions describes the physics in the length scale of $2nc_0$ of the hopping range.  Since the wavelength of the charge modulation is in the $c_0$ scale, a natural truncation  in the expansion has to perform to obtain correct physics.

The current-current interaction is of much smaller order of magnitude than the density-density interaction.  From the continuity equation, the order of magnitude of the current can be estimated by
\begin{equation}
J \sim \frac{\omega}{k}\rho \sim \frac{k_BT_{\text{CDW}}}{M_0}\rho,
\end{equation}
which is roughly $10^{-4}\rho$.  The current-current interaction is due to the Lorentz force of the spin-Berry's phase.   Performing a similar procedure and taking $\vec{J}(x)= \vec{J}_{\uparrow}(x)+\vec{J}_{\downarrow}(x)$,
\begin{eqnarray}
\frac{-ig^2}{2}\vec{J}(x)\cdot\frac{i}{k^2-M^2_1+i\eta}\vec{J}(x)\approx\frac{g^2}{2M^2_1}[(\frac{4}{M^2_1c^2_0}-1)J^2(x_i)-\frac{1}{M^2_1c_0^2}\sum_{j=i\pm\hat{e}_x,\atop i\pm\hat{e}_y }\vec{J}(x_i)\cdot\vec{J}(x_j)], \label{e4}
\end{eqnarray} 
where $M_1=M_0$ for the cuprates. 

In the two-dimensional motion, translational motion is equivalent to the rotational motion.  Due to the density-density interaction in Eq.~(\ref{e3}), the checkerboard density modulation implies that electrons are confined in the unit cell.  Therefore, length scale of the current is also in the scale of $c_0$.  The orbital rotational motion, described by the angular momentum $\vec{L}(x_i)$, can be estimated by $|\vec{L}| \sim c_0|\vec{J}|$.  Then, the first term in Eq.~(\ref{e4}) is proportional to $L^2(x_i)$, and the second term is proportional to $\vec{L}(x_i)\cdot\vec{L}(x_j)$.  The minus sign indicates that the orbital rotational motion favors the ferromagnetic configuration.  In the checkerboard modulation, the virtual pair-hopping process favors the next-nearest-neighbor interaction to be ferromagnetic as well.   As the lowest rotational energy state is given by $L_{\uparrow}=L_{\downarrow}=1$, the electron pair in each unit of the checkerboard has angular momentum $2$ or $0$.  Namely, the pair wavefunction can be of $s$-wave or $d$-wave.  The ferromagnetic orbital rotational motion represents the weak ferromagnetism in the Polar Kerr rotation experiments~\cite{kapitulnik2008, Lee2018}.

\vspace{0.5cm}
\noindent
\emph{The effective model}

\begin{figure}[htb]
\includegraphics[width=0.45\textwidth]{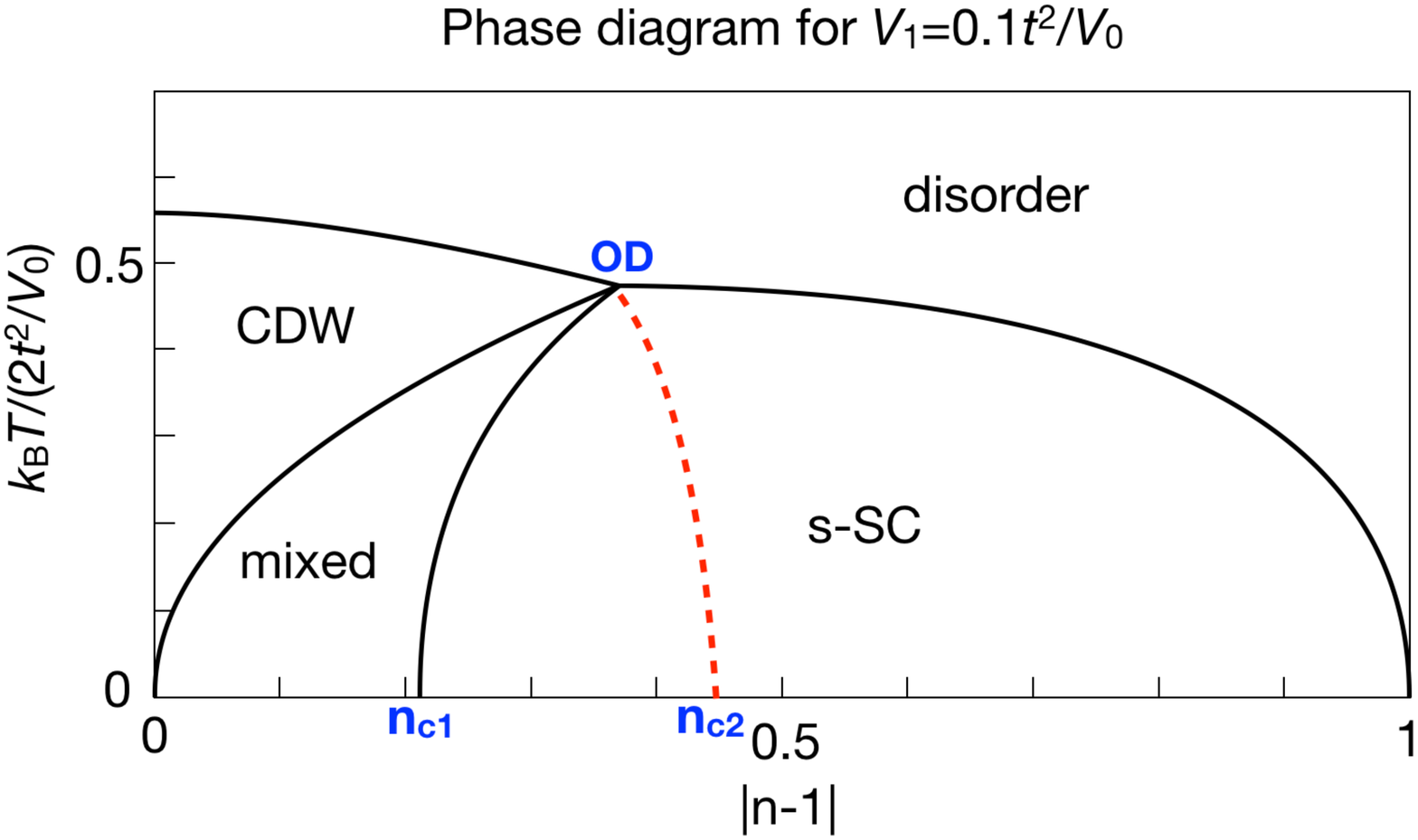}
\caption{The finite temperature phase diagram of the extended negative Hubbard model in the strong-coupling limit, that is $t < V_0$.  The vertical axis is the temperature, and the horizontal axis is the electron density ranging from $0 \le n \le 2$.  There are four phases: charge density wave, $s$-wave superconductivity, their mixed state, and the disorder phase.  The superconducting dome contains the mixed and the pure states.  The red dot line indicate that the superconducting dome should be truncated in cuprates, since $V_0$ and $V_1$ should be zero at 27\% doping.}\label{phase}
\end{figure}

Inspired by Eq.~(\ref{lagrangian2}), we suggest that the effective lattice models in the strongly correlated electron systems have two stages.  In the uniform state, the positive $U$ Hubbard model effectively describe the pseudogap formation.  On the occurrence of the quantum fluctuations of the correlation degrees of freedom, electronic structure becomes inhomogeneous.  The effective model is modified to be
\begin{eqnarray}
H_{\text{eff}}=-t\sum_{<ij>,\sigma}(c^\dagger_{i,\sigma}c_{j,\sigma}+\text{h.c.})-V_0\sum_{i}n_{i,\uparrow}n_{i,\downarrow} +V_1\sum_{<ij>,\sigma\sigma'}n_{i,\sigma}n_{j,\sigma'}, \label{LCmodel}
\end{eqnarray}
where $<ij>$ denote the nearest neighbors, and both $V_0$ and $V_1$ are positive.  Eq.~(\ref{LCmodel}) is nothing but the extended negative Hubbard model.  In addition, the current-current interaction also plays an important role.  The phase diagrams of Eq.~(\ref{LCmodel}) have been studied in the strong-coupling limit and the weak-coupling limit~\cite{chao1981, chao1982}.  In Fig.~(\ref{phase}), their results of the finite temperature phase diagram versus electron density ranging from $0 \le n \le 2$ in the strong-coupling limit is quoted for $V_1=0.1\frac{t^2}{V_0}$~\cite{chao1981}.  Since the quantum fluctuations is doing dependent,  we expect that $V_0$ and $V_1$ in Eq.~(\ref{LCmodel}) is doping dependent, as well.  The relevancy of the Eq.~(\ref{LCmodel}) to cuprates can be understood as the following.  Except for $V_1=0$, there are four phases: a charge density wave, the $s$-wave superconductivity, the mixed state of the above two, a disorder phase.  All phases are separated by a second order phase transition.  There is a superconducting dome containing two phases, pure and the mixed states~\cite{shen2014}.  The charge density wave occupies the finite temperature region vanishing at the optimal doping of the dome.  The mixed state naturally explains the recently-observed the pair-density-wave~\cite{julien2011, davis2018}, which vanishes at the doping $n_{c1}$.  The superconductivity in cuprates vanishes at around 27\% doping, suggesting that $V_0$ and $V_1$ should be zero at that doping.  Therefore, there should be another $n_{c2}$ truncating the superconducting dome for the real situation.

The extended negative Hubbard model hosts the $s$-wave superconductivity only.  It is not the complete effective model, since the current-current interaction is not included.  The current-current interaction, proportional to $-\vec{L}(x_i)\cdot\vec{L}(x_j)$, suggests an intra-unit-cell orbital motion~\cite{Lee2018}, in favor of a finite orbital angular momentum state illustrating the weak ferromagnetism.  The orbital angular momentum supplies the $d$-wave symmetry of the electron pairs.  The current-current interaction is small comparing to the density-density interaction.  The phase diagram remains the same subject to the $s$-wave superconductivity replaced by the $d$-wave superconductivity.

In the current approach, the effective electronic model is derived by integrating over the correlation degrees of freedom that are described by a non-compact massive U(1) gauge theory.  The compact version, namely the lattice counterpart, is not trivial.  A n\"aive construction of the compact abelian lattice gauge theory with the Higgs of the fixed magnitude was considered by Fradkin and Shenker~\cite{fradkin1979}.  If the Higgs field is not in the fundamental representation, which is our case, there are two phases separated by a phase boundary: confined phase and the Higgs phase.  In both phases, gauge field are massive.  Fradkin and Shenker commented that it is not trivial to find the continuous limit of their abelian lattice gauge theory~\cite{fradkin1979}.  Similar difficulty happens here.  Besides the detail mapping of the Eq.~(\ref{LCmodel}) to the cuprates, a large scale calculation to extract the doping dependence of $V_0$ and $V_1$, one of the future directions is to find a correspondent lattice gauge theory of which the current theory is the continuous limit.

\vspace{0.5cm}
\noindent
\emph{Summary}

In summary, our approach serves as a new recipe to study the correlated electrons.  The effect of the wavefunction overlaps delicately handcrafts new dynamics that organizes the correlated electrons to form a new state of matter in distinction to the Fermi liquid.  However, while the effective-model approach has been successful as the standard language, there are obstacles in the application to strongly correlated electron systems, in particular cuprates.  Firstly, the density modulation and the weak ferromagnetism arise without phase transitions.  \emph{They are actually not states but phenomena}~\cite{Lee2018}.  In the effective-model approach, they are described by physical states, and the problem to avoid phase transitions is inevitable.  Secondly, in the polar Kerr rotation experiment, the direction of the ferromagnetic moment is independent of the direction of the training field, which is intractable using the current approach.  

Identifying the correlation as the emergent degrees of freedom, the quantum fluctuations of the correlations give rise to diverse phenomena, including the instability to superconductivity~\cite{chern2018}.  Due to the large mass scale ($\sim$ 123 eV), the quantum fluctuations occur above the boiling point of the liquid nitrogen.  Although the occurrence of the quantum fluctuations vetoes to use a single effective model to describe all phenomena in the correlated electron systems, this new physical degrees of freedom have demonstrated many interesting existing properties, and there may be more to explore.

%\begin{figure}[ht
%\includegraphics[width=0.5\textwidth]{diffraction.pdf}
%\includegraphics[width=0.5\textwidth]{phase.pdf}
%\caption{(Color online) (A) The cartoon picture for the string excitation.  Flipping spin at site $i_1$ generates a monopole (green ball) and anti-monopole (blue ball) pair.  Flipping further at site $i_2$ makes the anti-monopole to hop.  In this figure, the anti-monopole hops between 5 tetrahedra following the blue dash line, which is the flux-tube excitation of the energy $\sim l\frac{K^2}{J}$.  The line in brown is an example of the pyrochlore hexagon which is the shortest path to annihilate the monopole and the anti-monopole.  (B) The phase diagram of the antiferromagnetic quantum Ising model in the pyrochlore lattice at zero longitudinal magnetic field.  The deconfinement only occurs at $K=0$.  With non-vanishing $K$, the system becomes the quantum confined phase and adiabatically connects to the paramagnetic phase.}\label{Fig:string}
%\end{figure}

\vspace{0.5cm}
\noindent
{\it Acknowledgement}

This work is supported by MOST 106-2112-M-002-007-MY3 of Taiwan.

%\bibliography{lattice}


\end{document}